\documentclass[USenglish,twocolumn]{article}

\usepackage[utf8]{inputenc}
\usepackage[big]{dgruyter}
 
\usepackage{graphicx}
\usepackage{txfonts}
\usepackage{isotope}
\usepackage{color}
\usepackage{setspace}
\usepackage{pdflscape}
\usepackage{float}

\def\mnras{MNRAS}
              
\def\aap{A\&A}
\def\aaps{A\&AS}
\def\apj{ApJ}
\def\apjs{ApJS}
\def\araa{ARA\&A}

\begin{document}

\articletype{Research Article{\hfill}Open Access}

\author*[1]{P\'eter N\'emeth}


\affil[1]{Astroserver.org, 8533 Malomsok, Hungary, E-mail: peter.nemeth@astroserver.org}


\title{\huge Atomic diffusion in the atmosphere of Feige\,86}
\runningtitle{Feige\,86 -- A forbidden fruit}


\begin{abstract}
{We have revisited the ultraviolet and optical spectra of the blue horizontal branch star Feige\,86. 
The new analysis finds the star cooler and more compact than previously determined.
The IUE spectrum of Feige\,86 holds numerous unidentified spectral lines of heavy metals, indicating efficient atomic diffusion in the atmosphere. 
Because diffusion plays a key role in the atmospheres of hot subdwarfs as well, it is indispensable to a better understanding of subdwarf pulsations and evolution.
Feige\,86 looks like an ideal target to confront diffusion theory with observations and test spectroscopic techniques.
Therefore, to advance our general understanding of diffusion in stellar atmospheres we urge for new ultraviolet spectroscopy of Feige\,86 at the highest possible resolution with HST/STIS.}
\end{abstract}

  \journalname{Open Astronomy}
\DOI{DOI}
  \startpage{1}
  \received{..}
  \revised{..}
  \accepted{..}

  \journalyear{2017}
  \journalvolume{1}

\maketitle
\section{Introduction}
Stars along the horizontal branch (HB) show a bimodal distribution. 
At the red (cool) end massive red-clump stars, that are close to the giant branch, dominate the observed distribution. 
On the blue (hot) end subdwarf B (sdB) stars dominate. 
The region in between is less densely populated and notable gaps exist in between the two extremes. 
Feige\,86 is near the cool end of the blue HB, therefore it is closely related to sdB stars. 
In fact, we assume that the same "canonical mass" helium burning core supports blue HB (BHB) stars like sdB stars. The major difference is the more massive envelope of BHB stars.  

Globular cluster studies (Brown et al., these proceedings) revealed that there is a great variety of HB morphologies in color-magnitude diagrams, almost as many different sequences as clusters. 
This diversity is not fully understood. The phenomena behind the different appearance of the HB were often referred to as the mysterious "second parameter". 
It can be metallicity, age or even a process such as atomic diffusion.
In sdB stars, and HB stars hotter than $\sim$11500 K diffusion is active and manifests itself, e.g. by the depleted helium abundance.
Still, a reassuring correlation between surface parameters from stratified atmosphere models in diffusive equilibrium and the morphology of the HB is not established. 
The reason is simple: such atmosphere models are quite complex and the analysis would also require a population study.  

However, it may well be that time has come to address diffusion in subdwarf atmospheres, and start applying stratified models.
Recent analyses of long orbital period binary stars revealed (Vos et al., Moni-Bidin et al. these proceedings) that sdB stars show a wider mass distribution than the canonical formation predicts. 
At the same time a notable spread of the EHB in the $T_{\rm eff}-\log{g}$ plane is observed, which is now significant over the statistical errors of atmospheric parameters. 
Even though the current models of stellar evolution and asteroseismology agree quite well on a canonical mass (e.g. $0.40-0.52 M_\odot$; \citealt{fontaine12}), there must be something we do not know well enough, such as:
\begin{itemize}
\item Current seismic models do not account properly for core convective overshooting.
\item Missing opacities and limited nuclear networks in evolution models.
\item Atmospheric parameters are affected by atomic diffusion and homogeneous models provide misleading boundary conditions.
\end{itemize}
All these are related to the observed mass discrepancies to a certain degree, and there may be further sources of uncertainties, such as the population membership of the stars analyzed. 

Atomic diffusion is an interesting process from a spectroscopist's view as it puts the core of model atmosphere analyses to a test. 
It is certainly at work in sdB stars and shapes their general abundance pattern: helium and light metals are depleted, while iron is close to the solar abundance and heavy metals show extremely high abundances (\citealt{naslim13}, \citealt{geier13}).  

For detailed investigations of diffusion high resolution UV spectroscopy is needed. 
However, there are only very few hot subdwarf stars bright enough to achieve a reasonable signal-to-noise ratio (SNR) with the currently available instruments. 
Chemically peculiar (CP) stars, in which the effects of diffusion are observable in the optical spectral range, are much cooler and fainter in the UV. 

Feige\,86 is a relatively bright (B=9.89 mag) star in the constellation Canes Venatici. 
It is classified as a BHB star and located between CP and sdB stars in the Hertzsprung–Russell diagram. 
Although Feige\,86 is well known for its peculiar helium lines for decades, it is still a poorly understood object on its own.
The optical helium line profiles show clear signs of chemical stratification and vertical isotopic separation of elements due to atomic diffusion.
Hence Feige\,86 challenges the current atmosphere studies.
However, these challenges also mark the direction for future improvements. 

\begin{figure}
\includegraphics[width=\linewidth]{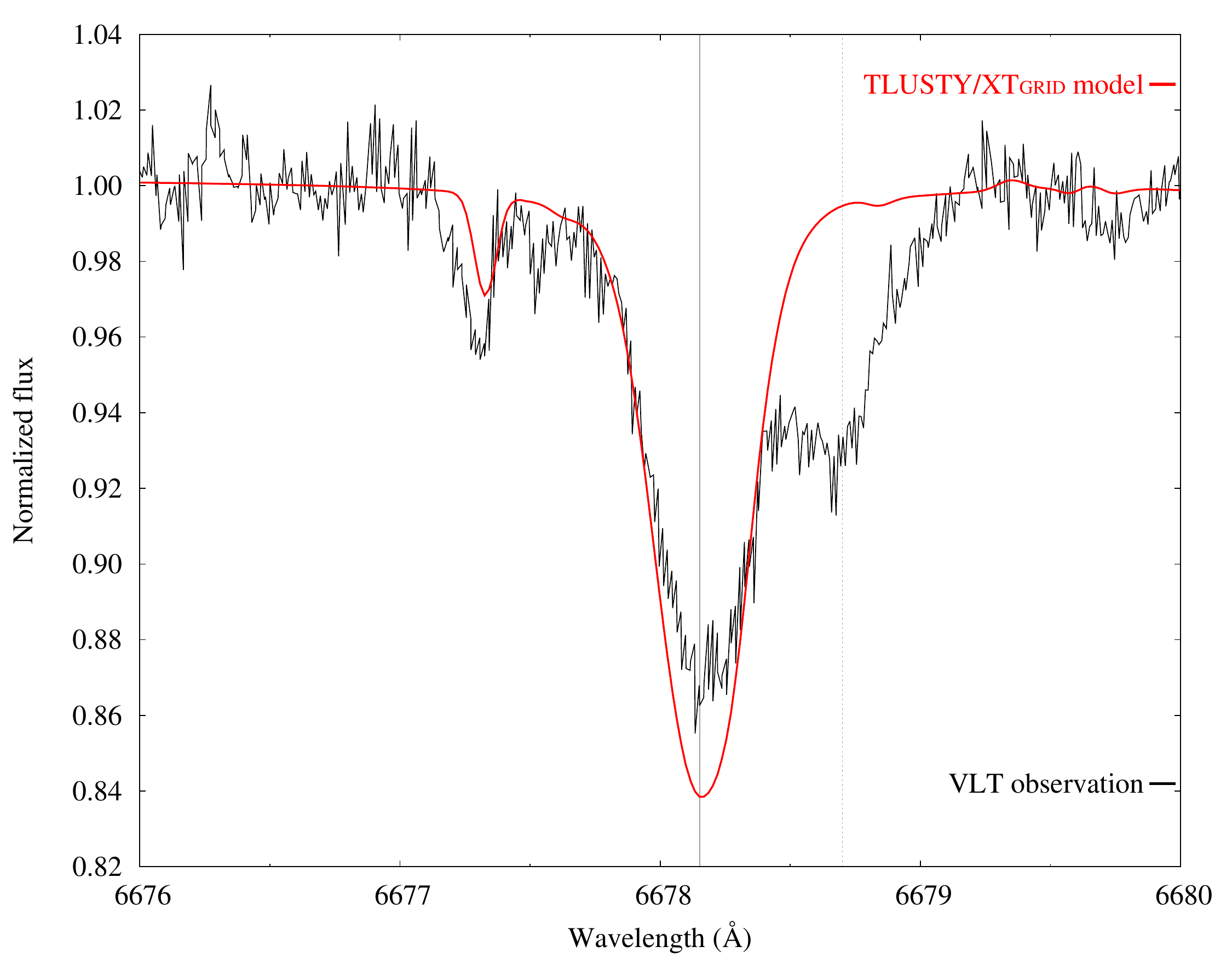}
\caption{Isotopic shifts in the optical spectrum of Feige\,86 are most remarkable in the helium lines. 
The He\,{\sc i} 6678 \AA\ line shows the isotopic separation of $^3$He (dashed line) and $^4$He (solid line). 
Vertical stratification causes the line profiles appear too strong in the wing and too shallow in the core when compared to homogeneous models. 
Isotopic shift broadens the lines and in the most extreme cases, such as the He\,{\sc i} {6678} \AA\ line, it separates the line into components.
\label{fig3}}
\end{figure}

Feige\,86 has a rich, sharp-lined peculiar
spectrum both in the ultraviolet (UV) and the optical. 
However, even for a peculiar star, its spectrum is quite unusual. 
The strong UV resonance lines of C and Si that
are present in B and sdB stars are all weak in Feige\,86, while metals, in
particular heavy metals show high abundances. 
The optical helium line
profiles indicate a strong helium isotopic shift and vertical stratification
(\citealt{hartoog79}, \citealt{cowley09}). 
In the optical range, the effects are strongest at the He\,{\sc i} 6678 \AA\ line. Figure \ref{fig3} compares the combined profile of $^3$He and $^4$He in the UVES observation to a homogeneous model.
All these properties coincide with the
predictions of atomic diffusion theory and make Feige\,86 a key target to
confront theory (\citealt{michaud1970}, \citealt{michaud11}) with observations.
Then, the methods developed for and tested on Feige\,86 can be extended to sdB stars. 

\section{Reanalysis with {\sc Tlusty/XTgrid}}
We consider Feige\,86 to be representative for peculiar BHB-tip stars and chose it as a benchmark object to streamline the development of our general fitting procedure ({\sc XTgrid}, \citealt{nemeth12}).
Therefore we use it here as a technical demonstration. 
{\sc XTgrid} uses {\sc Tlusty/Synspec} non-Local Thermodynamic Equilibrium (non-LTE) atmosphere models calculated in hydrostatic and radiative equilibrium (\citealt{hubeny17}) to iteratively fit observed spectra. 
The {\sc Tlusty} models treat chemical stratification at the level of vertical discretization of the atmospheric structure in a very flexible way. 
The problem arises in the actual fitting procedure as many new free parameters appear, which destroy all the preferred good properties of the $\chi^2$ landscape, practically making it flat. 
Even after adding extra constraints, such as a correlated variation of adjacent depth abundances for each element to make the chemical profiles smooth, or using theoretical predictions for the abundance gradients with depth, the convergence rate is low.
We consider a modified version of the genetic algorithm (e.g. \citealt{charbonneau95}) as the only practical minimization procedure to work with. 
We keep the global $\chi^2$ minimization procedure to fit the global parameters, such as temperature, gravity and metallicity.
The genetic approach is applied only to the vertical redistribution of the individual abundances. 
Tests show this approach to be sufficient for helium and promising for abundant elements such as iron.

To address the vertical distribution of
elements, first, one must determine the global metallicity profile (the individual abundances) and
construct a line list optimized for the target type. 
This is a standard procedure for normal stars, but when diffusion is present the situation is more complex. 
The parameter determination depends on the signal-to-noise ratio of the observations, the quality of atomic data, and the unknown effects of diffusion.
In each cases, either
only the line profiles, or the equivalent width ratios of narrow metal lines
can give information on the vertical element stratification. 
Such work requires high quality observations as well as reliable atomic data. 
For our {\sc XTgrid} fit, we took the R=10\,000
IUE/SWP ultraviolet spectrum of Feige\,86 with a mean SNR of $\sim$25 and complemented this data with VLT/UVES spectra from the ESO
Archive with R=71\,000-107\,000 and SNR of 30-80. 

In our first attempt, we
performed a simultaneous global fit of the two data sets to determine the
mean atmospheric parameters, find the most abundant species, and refine
the atomic data of their strongest lines. 
The global approach on the UV-optical observations assures that all relevant transitions from the line list are included in the model. 
The method also exploits the combined constrains provided by the different data sets. 
The optical range is ideal to constrain $T_{\rm eff}$, $\log{g}$, the helium abundance and the projected rotation velocity from the Balmer and the helium line profiles and strengths. 
For metallicity, the UV range represents a stronger constraint through line-blanketing and also extends the range over which the ionization equilibria are examined.

\begin{figure*}
\includegraphics[width=\textwidth]{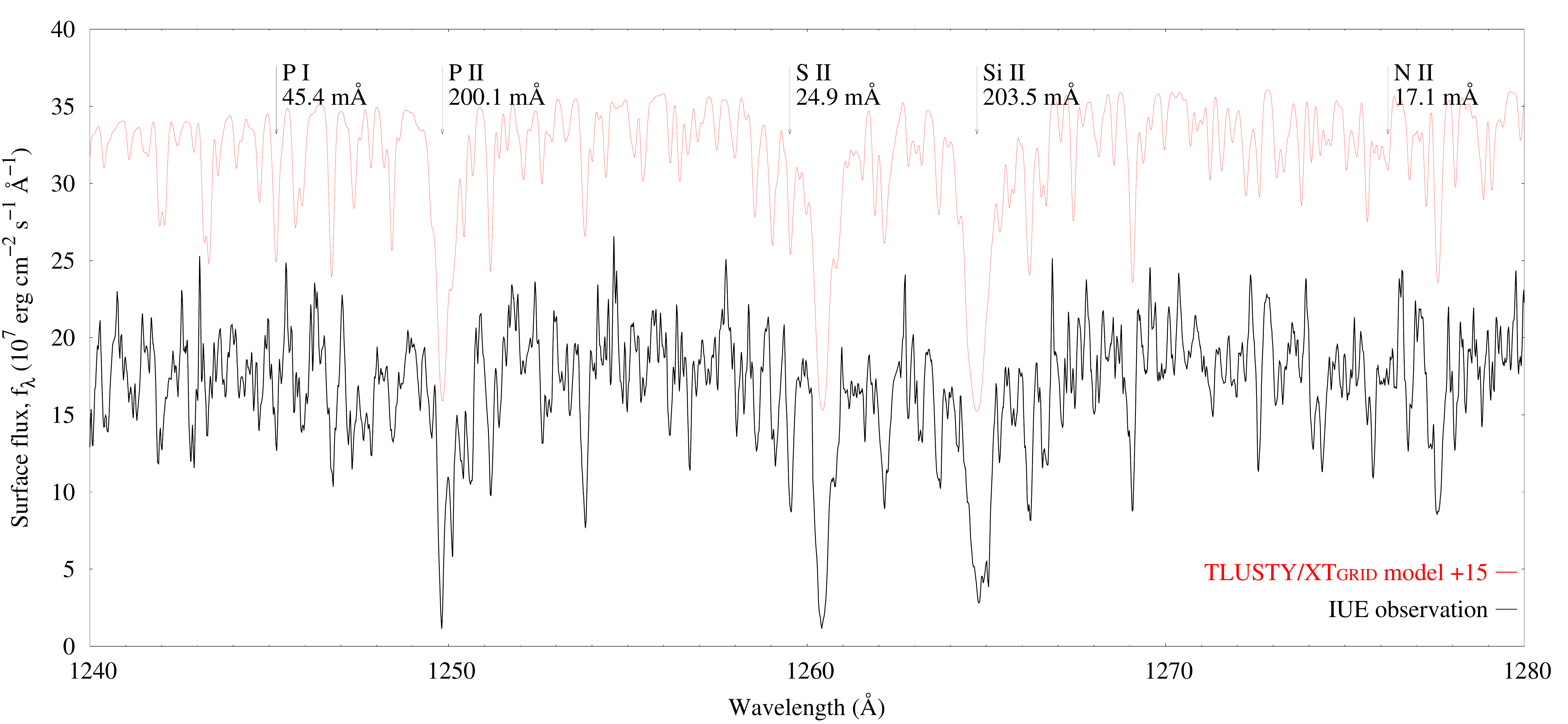}
\caption{Part of the IUE/SWP observation of Feige\,86 reveals a very crowded spectrum with a wealth of heavy metal lines.
However, the characteristic strong resonance lines of light metals are all remarkably weak. 
The strongest spectral lines that have an equivalent width larger than 10 m\AA\ in the 1240-1280 \AA\ spectral range are marked for each ion.
The strong lines of P correspond to a $\sim$200 times solar abundance, which is another indication of active atomic diffusion in Feige\,86. 
\newline Note: We fitted the entire UV-optical range with a single model in search for a global best-fit, which explains some of the discrepancies seen here, when a small spectral range is drawn from the global fit.
\label{fig1}}
\end{figure*}

\begin{figure*}
\includegraphics[width=\textwidth]{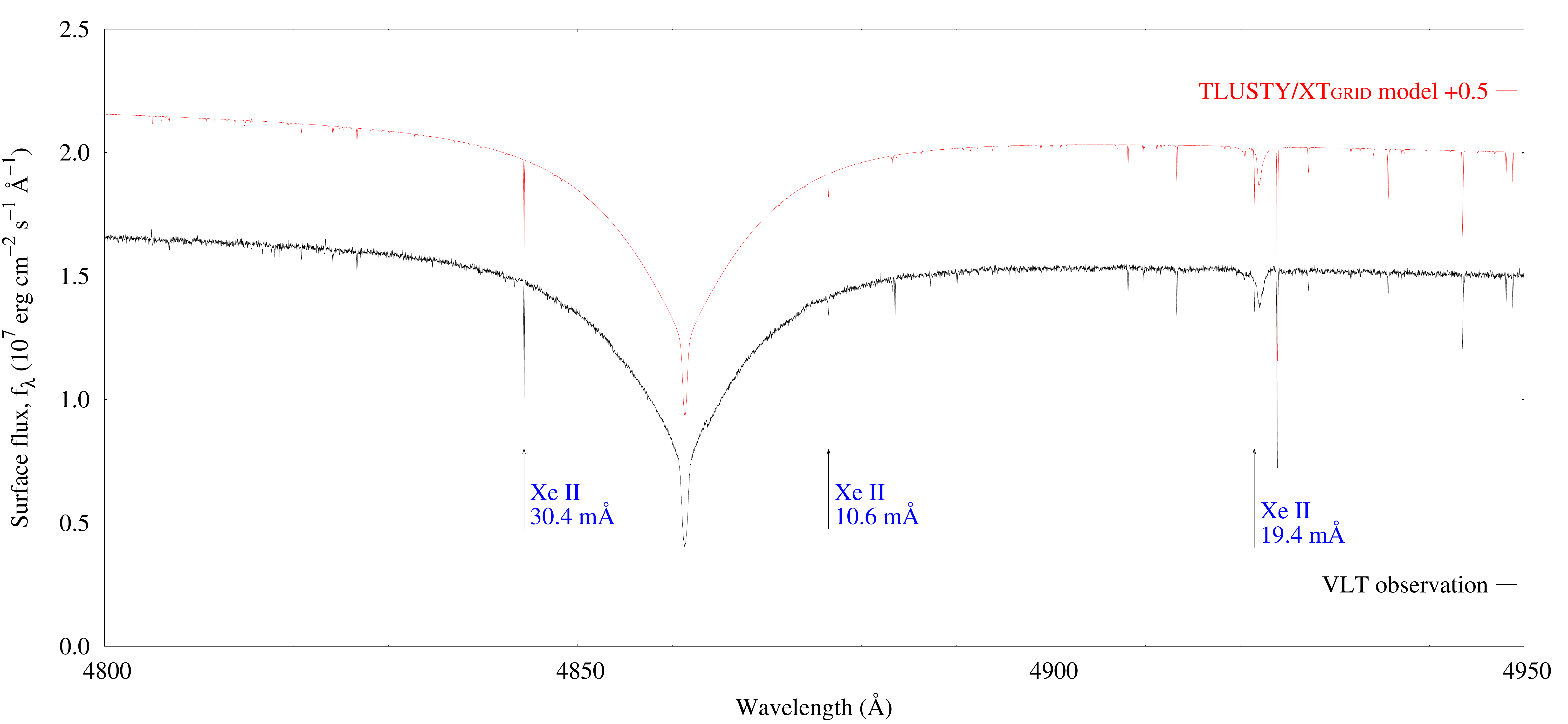}
\caption{ The VLT/UVES observation of Feige\,86 reveals strong lines of Xe\,{\sc ii} in the optical. 
The three spectral lines of Xe\,{\sc ii} that have an equivalent width larger than 10 m\AA\ in the 4800-4950 \AA\ spectral range are marked with blue labels. All lines of Xe are marked in the online version.
\newline Note: We fitted the entire UV-optical range with a single model in search for a global best-fit, which explains some of the discrepancies seen here, when a small spectral range is drawn from the global fit.
\label{fig2}}
\end{figure*}

To our surprise, we found a set of
strong heavy metal lines in the optical spectrum, all corresponding to Xe\,{\sc ii}
lines and a nearly 20\,000 times solar Xe abundance. 
Some of the Xe\,{\sc ii} lines
(e.g. $\lambda$4180, $\lambda$4208, $\lambda$4245, $\lambda$4330, $\lambda$4462 \AA) show a $\sim{0.1}$ \AA\ blue shift, which indicates that we see different isotopes at once, similar to the case of helium. 
The radial velocity shift of the lines corresponds to the isotopic shifts observed in CP stars \citep{yuce11}.
Past studies have reported Xe, Bi and other heavy elements in Feige\,86 (\citealt{bonifacio95}, \citealt{castelli97}). 
Therefore, we extended the spectral synthesis capabilities of {\sc XTgrid} to include atomic
data for the whole periodic table and included 78 elements in this work. 
The strongest lines of each significant element and all strong lines of Xe\,{\sc ii}
are labeled in our {\sc XTgrid} fit online. 
Figure \ref{fig1} and \ref{fig2} show small, but representative parts of the fit to the UV and optical spectra, respectively.

The presence of Xe\,{\sc ii} and heavy metals in general is not surprising in CP stars. 
Recent studies of HgMn stars found that, where Xe was detected, the mean Xe abundance was much higher than the cosmic abundance (\citealt{dworetsky08}, \citealt{cowley10}). 
On the other hand, these studies found no signs of Xe absorption features in normal, late B type stars.
Therefore, the presence of Xe may be a clear indication for diffusion in B star atmospheres. 
Fortunately, the $\lambda$4844.33 \AA\ Xe\,{\sc ii} line is strong enough to be detectable even with low resolution spectroscopy.

Compared to previous analyses, the new effective temperature and surface gravity of Feige\,86 are quite different. 
\cite{bonifacio95} have found $T_{\rm eff}=16430\pm250$ K and $\log{g}=4.2\pm0.02$ cm\,s$^{-2}$ using LTE {\sc Atlas-9} models, while we found 
$T_{\rm eff}=15025\pm275$ K and $\log{g}=4.56\pm0.06$ cm\,s$^{-2}$ with non-LTE {\sc Tlusty} models. 
The Eddington luminosity fractions are $\log{L/L_{\rm Edd}}=-2.46$ and $\log{L/L_{\rm Edd}}=-2.97$, corresponding to a luminosity fraction ratio of 3. 
The value of $\log{L/L_{\rm Edd}}=-2.97$ also corresponds to the zero-age extreme HB.
This reflects a substantial difference between the results, which is expected as both the models and the observational data are different, and more than 20 years have passed in between the two analyses.
\citet{maza14} also found significant non-LTE effects in the atmospheres of HgMn stars, which play a role both in the abundance determinations as well as in the determination of the vertical stratification of elements.

\section{Flash or intershell nucleosynthesis?}

\begin{figure*}
\includegraphics[width=\textwidth]{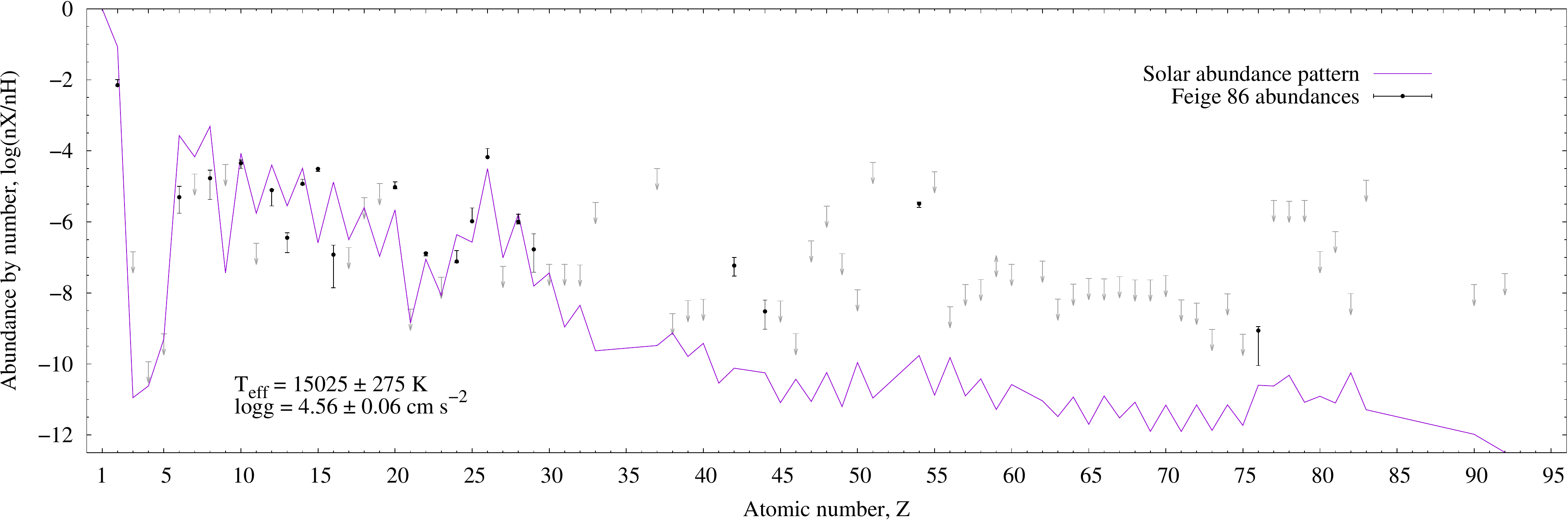}
\caption{The abundance pattern of Feige\,86, compared to the solar mixture {\citep{asplund09}}. 
While most of the heavy metal abundances are only upper limits, the overabundance of Xe hints the presence of a hitherto unseen peak of heavy elements.\label{fig4}}
\end{figure*}

The extreme abundance of Xe and its nine naturally occurring isotopes make
it a good trace element to search for the nucleosynthesis history of Feige\,86. 
Heavy metal formation requires a high flux of neutrons. 
The observed Xe
may be primordial or may have formed by the s-process in the intershell region of the progenitor giant star during subsequent shell He-flashes. 

A further
ambiguity arises from diffusion, which largely changes the observable
abundance pattern (Figure \ref{fig4}). 
Therefore, while the presence of Xe is
confirmed in the atmosphere of Feige\,86, finding its origin will require future work. 
The process that produced Xe also produced further trans-iron elements, which are
brought to the atmosphere by dredge-up and pushed to the surface by
diffusion.
To search for these signatures, new and higher resolution UV spectra are necessary. 
Fortunately, thanks to the brightness of Feige\,86, such data can be obtained with relatively short visits of the Hubble Space Telescope. 


\section{Summary}

Twenty years after the comprehensive atmospheric analyses of \citet{bonifacio95} and \citet{castelli97} we revisited Feige\,86 with new models applied on new observational data. 
According to our results, the star is somewhat cooler and more compact than previously determined. 
The Xe\,{\sc ii} 4844.33 \AA\ line turned out to be an excellent diagnostics for active diffusion in BHB stars.

We found Feige\,86 to be a particularly interesting BHB star. 
Its effective temperature $T_{\rm eff}\approx15\,000$ K is well above the $\sim$11\,500 K boundary where convective turbulence is inefficient to erase the effects of atomic diffusion. 
The metallicity profile is very similar to CP stars, in particular to HgMn stars on the main-sequence, while the  surface gravity of Feige\,86 is at least 0.5 dex larger than for normal CP stars.
Therefore Feige\,86 is a compact star between HB and sdB stars.
No other BHB-tip star is known to us brighter than V=10 magnitude.
Feige\,86 offers a unique opportunity to investigate HB stars with atomic diffusion. 
Thanks to the similar characteristics this work can be done in parallel with CP stars and eventually extended to sdB stars. 

The brightness and the complexity of the atmosphere of Feige\,86 make it a significant target to learn
about hot atmospheres, NLTE conditions and radiative transfer, atomic data,
diffusion as well as stellar nucleosynthesis. 
To understand this complexity, high-resolution FUV and NUV spectroscopy is needed together with a sophisticated spectral analysis procedure, that is now under development.

\vspace{5pt}
\noindent More information on Feige\,86 and the fit made to the entire range of the IUE-SWP and VLT-UVES spectra are available at:
\newline\centerline{\url{www.astroserver.org/asyyl4}}


\end{document}